\documentclass{aa}  

\usepackage[varg]{txfonts}
\usepackage{multirow}
\usepackage{graphicx}
\usepackage{color}
\usepackage[colorlinks, citecolor={blue}]{hyperref}
\bibpunct[; ]{(}{)}{;}{a}{}{;}

\titlerunning{The mass and environmental dependence on the secular processes of AGN}

\begin{document}

\title{The mass and environmental dependence on the secular processes of AGN in terms of morphology, colour, and specific star-formation rate}

\author{
M.~Argudo-Fern\'andez\inst{1,2}
\and
I.~Lacerna\inst{3,4,5}
\and 
S. Duarte Puertas\inst{6}
}

\institute{
Centro de Astronom\'ia (CITEVA), Universidad de Antofagasta, Avenida Angamos 601 Antofagasta, Chile
\email{maria.argudo@uantof.cl}
\and 
Chinese Academy of Sciences South America Center for Astronomy, China-Chile Joint Center for Astronomy, Camino El Observatorio, 1515, Las Condes, Santiago, Chile
\and 
Instituto de Astronom\'ia, Universidad Cat\'olica del Norte, Av. Angamos 0610, Antofagasta, Chile
\and
Instituto Milenio de Astrof\'isica, Av. Vicu\~na Mackenna 4860, Macul, Santiago, Chile
\and
Instituto de Astrof\'isica, Pontificia Universidad Cat\'olica de Chile, Av. V. Mackenna 4860, Santiago, Chile 
\and         
Instituto de Astrof\'isica de Andaluc\'ia (CSIC) Apdo. 3004, 18080 Granada, Spain}

   \date{Received 2 May 2018; accepted 3 October 2018}

 
\abstract
{Galaxy mass and environment play a major role in the evolution of galaxies. In the transition from star-forming to quenched galaxies, Active galactic nuclei (AGN) have also a principal action. However, the connections between these three actors are still uncertain.}
{In this work we investigate the effects of stellar mass and the large-scale environment (LSS), on the fraction of optical nuclear activity in a population of isolated galaxies, where AGN would not be triggered by recent galaxy interactions or mergers.}
{As a continuation of a previous work, we focus on isolated galaxies to study the effect of stellar mass and the LSS in terms of morphology (early- and late-type), colour (red and blue), and specific star formation rate (quenched and star-forming). To explore where AGN activity is affected by the LSS we fix the stellar mass into low- and high-mass galaxies. We use the tidal strength parameter to quantify their effects.}
{We found that AGN is strongly affected by stellar mass in ``active'' galaxies (namely late-type, blue, and star-forming), however it has no influence for ''quiescent'' galaxies (namely early-type, red, and quenched), at least for masses down to $\rm 10^{10}\,[M_\odot]$. In relation to the LSS, we found an increment on the fraction of SFN with denser LSS in low-mass star forming and red isolated galaxies. Regarding AGN, we find a clear increment of the fraction of AGN with denser environment in quenched and red isolated galaxies, independently of the stellar mass.}
{AGN activity would be ``mass triggered'' in ``active'' isolated galaxies. This means that AGN is independent of the intrinsic property of the galaxies, but on its stellar mass. On the other hand, AGN would be ``environment triggered'' in ``quiescent'' isolated galaxies, where the fraction of AGN in terms of sSFR and colour increases from void regions to denser LSS, independently of its stellar mass.}


\maketitle

%

\section{Introduction}

Galaxy mass and environment are both playing a major role in the evolution of galaxies from star-forming disk to passive spheroidal galaxies. This evolution can happen in a short time scale, where the star formation activity is truncated by a discrete event, or could be due to a gradual increase of the average age of the stellar population \citep{2015MNRAS.451..888C}. Two independent mechanisms, one related to stellar mass and another related to environment, have been proposed to explain this transition, generally called as ``quenching'', for local galaxies \citep{2010ApJ...721..193P,2012ApJ...757....4P}. In this regard, there would be a ``environmental quenching'' driven by galactic over-density \citep[e. g.,][]{2014MNRAS.444.2938H,2016MNRAS.462.3955P,2017MNRAS.467.4200R}, and ``mass quenching'', where the shut down of star formation occurs on more dynamic processes \citep[e. g.,][]{2014MNRAS.441..599B,2016MNRAS.459..754F,2018MNRAS.473.1168R}. 

The  ``quenching'' process has been broadly studied using diagnostic diagrams as the star formation rate (SFR) versus stellar mass diagram, the specific star formation rate (sSFR) versus stellar mass diagram, or the colour-stellar mass \citep[e.g.,][]{2004MNRAS.351.1151B,2007ApJ...660L..43N,2011ApJ...739L..40R,2014MNRAS.440..889S,2015ApJ...801L..29R,2017A&A...599A..71D,2017ApJ...848...87C}. In the SFR-stellar mass diagram, for example, active star forming galaxies define the main sequence at high values of the SFR, where starburst galaxies produced by merger processes appears as a outliers galaxies to this main sequence \citep[e.g.,][]{2011ApJ...739L..40R,2015MNRAS.451..888C}. Conversely, the quenched galaxies have higher stellar masses and lower SFR values. The use of these diagrams to study galaxies in different environments shows, for instance, that field galaxies are mainly located in star forming regions (the so-called ``blue cloud''), whereas the group central galaxies are strongly biased to the passive regions (the so-called ``red sequence''), which can be explained by the fact that the stellar mass distributions of these two populations are different \citep{2014ApJ...788...29L}. The transition area between the two regions is commonly known as the ``green valley'' \citep[e.g.,][]{2014MNRAS.440..889S}. 

The principal mechanism responsible for ``quenching'' the star formation in galaxies is still unclear \citep[e.g.,][]{2015Natur.521..192P,2017MNRAS.471.2687B,2018MNRAS.476...12B,2018MNRAS.473.5617C,2018MNRAS.473.1346N,2018MNRAS.473.2679S}. However, observations have led to the interpretation that active galactic nuclei (AGN) is a plausible physical mechanism of the transformation of star-forming galaxies into passive galaxies \citep[e.g.,][]{2006MNRAS.370..645B,2014ARA&A..52..589H,2016Natur.533..504C,2016A&A...588A..78B,2016MNRAS.463.2986S,2018MNRAS.476..979P}. Accordingly, the green valley would be mainly populated by galaxies hosting an AGN \citep{2017ApJ...844..170T}. Unfortunately, the connections between AGN, stellar mass, and environment are still not well understood \citep[e.g.,][]{2017NatAs...1E.165H,2018arXiv180103884C}.

Major mergers appear to be the main driver of high luminosity AGN \citep{2011MNRAS.418.2043E,2012MNRAS.419..687R,2015MNRAS.452..774K,2014ApJ...788..140M,2014MNRAS.441.1297S,2015ApJ...804...34H,2015ApJ...806..147C}. This result would be connected to galaxy mass since the AGN-driven feedback has been usually proposed as a mechanism of quenching for massive galaxies after a major merger \citep[e.g.,][]{1998A&A...331L...1S}. Besides, AGN feedback is expected to play a mayor role in galaxies with massive halos \citep[e.g.,][]{2012MNRAS.424..190G}, typically $\rm M_h~\gtrsim~10^{12}\,[M_\odot]$ \citep{2006MNRAS.370..645B,2006MNRAS.365...11C}, which corresponds to approximately $\rm M_\star~\gtrsim~10^{10.5}\,[M_\odot]$. On the contrary, \citet{2013MNRAS.430..638S,2015MNRAS.447..110S} suggest that the large-scale environment and galaxy interactions play a fundamental but indirect role in AGN activity, by influencing the gas supply, where the dependence on AGN luminosity is minimal. 

The connection between environment and nuclear activity is not yet clear, however there is some consensus that secular processes may be much more important in driving the black hole growth than previously assumed \citep{2015aska.confE..83M,2015MNRAS.447..110S}, where an abundant supply of central cold gas, regardless of its origin, is the principal trigger of AGN, which can be fed by secular processes. Black holes growth by secular processes are dominant in isolated galaxies \citep{2008A&A...486...73S,2013MNRAS.433.1479H,2015MNRAS.451.1482M}. The purpose of this study is therefore to identify the interplay role of the stellar mass and the large-scale environment (LSS) on driving AGN activity. We refer to "mass triggered" or "environment triggered" when the AGN is mostly triggered by stellar mass or environment, respectively, in a population of isolated galaxies where AGN would not be triggered by recent galaxy interactions or mergers. With this work we are extending a previous study carried out by \citet{2016A&A...592A..30A}. They observed different trends depending on galaxy mass. The fraction of optical AGN for low-mass isolated galaxies decreases from voids to denser regions, meanwhile the fraction of optical AGN for high-mass isolated galaxies increases with denser large-scale environment. Since the prevalence of AGN also depends on galaxy morphology \citep{1995ApJ...438..604M,2010ApJ...714L.108S,2012A&A...538A..15H,2012A&A...545A..15S}, in this work we go one step further to explore these trends as a function of galaxy morphology, colour, and specific star formation rate. This work is complemented by a companion study \citep{Lacerna+2018}, where we focus on isolated elliptical galaxies to explore the connection of the large-scale environment in their integrated properties.

This study is organised as follows. In Sect.~\ref{Sec:data} we describe the sample of isolated galaxies used in this work as well as the selected AGN classification method, the morphology classification criteria to separate the sample into early- and late-type galaxies, the relations considered to separate galaxies between red/blue and star-forming/quenched, and the parameters used to quantify the environment. We present our results in Sect.~\ref{Sec:res} and the associated discussion in Sect.~\ref{Sec:dis}. Finally, a summary and the main findings of the study are presented in Sect.~\ref{Sec:con}. Throughout the study, a cosmology with $\Omega_{\Lambda 0} = 0.7$, $\Omega_{\rm{m} 0} = 0.3$, and $H_{0}=70$\,km\,s$^{-1}$\,Mpc$^{-1}$ is assumed.

\section{Data and methodology} \label{Sec:data}

\subsection{Isolated galaxies}

Isolated galaxies represent a population of galaxies with minimized environmental evolutionary effects. We use galaxies of the SDSS-based catalogue of isolated galaxies \citep[hereafter SIG;][]{2015A&A...578A.110A}. SIG galaxies are selected from the SDSS-DR10 \citep{2014ApJS..211...17A} in a volume limited sample redshift range $0.005 \leq z \leq 0.080$, with $11 \leq m_{r} \leq 15.7$, where $m_{r}$ is the SDSS model magnitude in the $r$-band. The SIG is composed of 3702 isolated galaxies, which represents about 11\% of the galaxies in the local Universe \citep[$z\leq0.080$;][]{2015A&A...578A.110A}. 

Under a three dimensional isolation criterion, SIG galaxies are isolated with no neighbours in a volume of 1\,Mpc projected distance within a line-of-sight velocity difference of $\Delta\,\varv~\leq~500$\,km\,s$^{-1}$, and at least 2 orders of magnitude fainter within the range of spectroscopic completeness of the SDSS main galaxy sample at $m_{r,\rm{Petrosian}}~<~17.77$\,mag \citep{2002AJ....124.1810S}. See \citet{2015A&A...578A.110A} for further details.

\subsection{Stellar masses and AGN classification} \label{Sec:AGN}

We used published stellar masses and AGN classifications for galaxies in the SDSS-DR7 from \citet{2013MNRAS.430..638S}, hereafter SBA13 classification. In particular, we focus on optically selected nuclear activity. 

\defcitealias{2013MNRAS.430..638S}{SBA13}

Total stellar masses and AGN classification in \citetalias{2013MNRAS.430..638S} were drawn from \citet{2003MNRAS.341...33K} and from BPT diagnostic \citep{1981PASP...93....5B,2003MNRAS.346.1055K}, respectively. The information about optical spectra, as the corrected emission-line fluxes necessary to built BPT diagrams, drawn from the Max Plank Institute for Astrophysics and Johns Hopkins University \citep[MPA-JHU\footnote{Available at \texttt{http://www.mpa-garching.mpg.de/SDSS/DR7/}};][]{2003MNRAS.341...33K,2004ApJ...613..898T,2007ApJS..173..267S} added value catalogue \citep{2004MNRAS.351.1151B}. 

\begin{figure}
\centering
\includegraphics[width=\columnwidth]{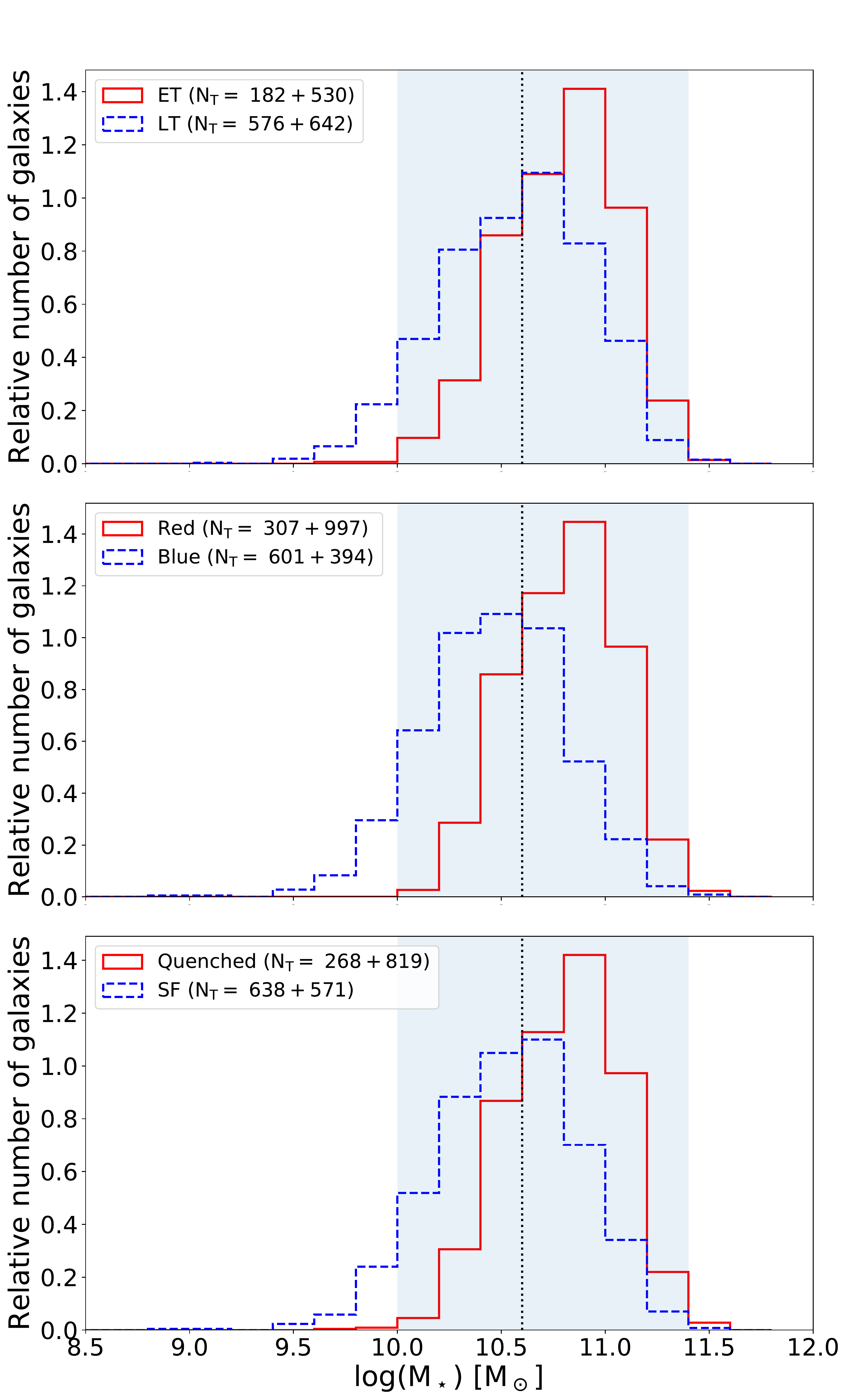}
\caption{Distribution of the stellar masses for galaxies in the SIG sample according to their morphology (red solid line for early-type galaxies and blue dashed line for late-type galaxies in the top panel), colour (red solid line for red galaxies and blue dashed line for blue galaxies in the middle panel), and specific star formation rate (red solid line for quenched galaxies and blue dashed line for star-forming galaxies in the lower panel). The shaded area in the figures corresponds to the selected stellar mass range for this study at $10.0~\leq~\rm{log}(M_\star)~\leq~11.4~[M_\odot]$. The vertical black dotted line corresponds to the selected mass cut to separate between low-mass and high-mass galaxies at $\rm{log}(M_\star)~=~10.6~[M_\odot]$. The total number of galaxies at each side of the mass cut is indicated in the corresponding legend.}
\label{Fig:mass}
\end{figure}

The fraction of AGN depends strongly on stellar mass \citep{2003MNRAS.346.1055K,2010ApJ...721..193P,2012ApJ...757....4P}. Therefore, to have enough numbers of galaxies in each mass bin, we considered galaxies with stellar masses within the range $10.0~\leq~\rm{log}(M_\star)~\leq~11.4~[M_\odot]$ (see the three panels in Fig.~\ref{Fig:mass}). There are 2299 SIG galaxies in this stellar mass range and with available \citetalias{2013MNRAS.430..638S} classification. The total number of these galaxies classified by morphology, colour, and specific star formation rate (sSFR) (when it is possible to classify, see next sub-sections) is shown in the first row of Table~\ref{tab:samples} (see Sections \ref{Sec:morpho} and \ref{Sec:color}). Note that there may be repeated galaxies in each classification column, for instance some SIG late-type galaxies can be also classified as blue galaxies. 

Following \citetalias{2013MNRAS.430..638S}, AGN classification includes transition objects (TO), Seyfert (Seyfert 1 not included), and low-ionization nuclear emission-line region \citep[LINER;][]{1980A&A....87..152H} galaxies. To have a statistically significant number of galaxies when dividing our sample by morphology, colour, and sSFR, we only separate nuclear activity into AGN, star-forming nuclei (SFN), and passive galaxies (in case that no nuclear activity is detected). To check that our results are not biased by this separation, we also explore the observed trends considering the different AGN subtypes in the Appendix \ref{Sec:AppB}. The number of galaxies classified in each type of nuclear activity is also shown in Table~\ref{tab:samples}. 

Note also that, for the reason above, we do not impose any limit on the $[\rm O_{III}]_{5007}$ emission line luminosity. We therefore need to take into account that some low luminosity AGN (usually LINERs) at higher redshifts could be classified as passive if their emission is not strong enough to be detected. Additionally, we discuss our results when considering a different AGN classification, composed of TO and Seyfert galaxies in Sect.~\ref{Sec:dis1}.

\begin{table}
 \caption{Number of galaxies in each subsample; ET: early-type galaxies as defined in Sect.~\ref{Sec:morpho}; LT: late-type galaxies as defined in Sect.~\ref{Sec:morpho}; red: red galaxies as defined in Sect.~\ref{Sec:color}; blue: blue galaxies as defined in Sect.~\ref{Sec:color}; quenched: centrally quenched galaxies as defined in Sect.~\ref{Sec:color}; SF: centrally star-forming galaxies as defined in Sect.~\ref{Sec:color}.}
 \label{tab:samples}
\centering 
\begin{tabular}{lcccccc}
\hline
 Type & ET & LT & red & blue & quenched & SF  \\
\hline
Total              & 712 & 1218 & 1304 & 995 & 1087 & 1209 \\
\\
AGN                & 362 & 586 & 768 & 385 & 611 & 541 \\
\,\,\,\,\, TO      & 121 & 332 & 299 & 249 & 138 & 409 \\
\,\,\,\,\, Seyfert & 37  & 65  & 77  & 46  & 62  & 60  \\
\,\,\,\,\, LINER   & 204 & 189 & 392 & 90  & 410 & 72  \\
SFN                & 44 & 462 &101 & 486 & 8 & 577 \\
Passive            & 306 & 170 & 435 & 124 & 468 & 91 \\
\end{tabular}
\tablefoot{Meaning of the different types. Total: total number of galaxies in each subsample; AGN: galaxies classified as LINER, Seyfert, or transition objects (TO); SFN: star-forming nuclei galaxies; passive: galaxies with no optical nuclear activity.}
\end{table}

\subsection{Morphology} \label{Sec:morpho} 

Galaxy morphology is another fundamental property to understand galaxy evolution. In the hierarchical galaxy formation scenario, major mergers is the prevailing mechanism of formation of early-type galaxies, which commonly happen in dense regions \citep[e.g.][]{Hernquist1993, Kauffmann1996, Tutukov+2007, 2014MNRAS.440..889S}. Thus, it is expected that a high fraction of the population of isolated galaxies is composed of late-type galaxies \citep{Dressler1980, Hernandez-Toledo+2010, 2014ApJ...788...29L,Vulcani+2015}. A previous analysis of AGN activity in isolated elliptical galaxies in the local Universe was performed in \citet{Lacerna+2016}. In this paper, we aim to investigate the AGN activity for both isolated early-type galaxies and isolated late-type galaxies.

We used the automated morphology classification for a representative sample of galaxies developed by \citet{2011A&A...525A.157H}. They provide a catalogue with a detailed, automated computational morphological classification for 698,420 galaxies in the SDSS-DR7 \citep{2009ApJS..182..543A} spectroscopic catalogue with redshift less than 0.25 and $m_{r} \leq 16$\,mag. \citet{2011A&A...525A.157H} divided the morphologies into four classes (E, S0, Sab, and Scd) and a broader early-type class (E or S0). Their automated method associates each galaxy with a probability to belong to each one of those categories. We considered the probability to be an early-type galaxy ($p_{\rm Early}$) to separate between early- and late-type galaxies.

There are 2293 SIG galaxies with morphological classification, in the stellar-mass range considered this study and with available \citetalias{2013MNRAS.430..638S} classification.
Of them there are 712 early-type galaxies (when $p_{\rm Early}\,>\,0.7$) and 1218 late-type galaxies (when $p_{\rm Early}\,<\,0.3$).

\subsection{Colour and specific star formation rate} \label{Sec:color}

We also aim to study the fraction of AGN, SFN, and passive galaxies as a function of environment in terms of galaxy properties such as colour and specific star formation rate (sSFR). We use the relation found by \citet{2014ApJ...788...29L} to separate red and blue galaxies, specifically

\begin{equation} \label{eq_gi}
(g - i)  = 0.16\,[\rm{log}(M_{\star})-10.31]+1.05  \quad,
\end{equation} 
where the the magnitudes $g$ and $i$ were taken from the SDSS database with extinction corrected $modelMag$ magnitudes  (\verb|dered| parameter), and the stellar mass is in units of M$_\odot$. We appended/added to our catalogues the photometric information/data from SDSS-DR12 \citep{2015ApJS..219...12A}.
With this method we were able to classify all the SIG galaxies in our study in terms of colour. The SIG galaxies lying below the relation in the $(g - i)$--$\rm{log}(M_\star)$ diagram are classified as blue (995 SIG galaxies), meanwhile SIG galaxies above the line are classified as red (1304 SIG galaxies). 

Likewise, we use the relation provided by  \citet{2014ApJ...788...29L} to separate quenched and star-forming galaxies in terms of their sSFR, i.e.,

\begin{equation} \label{eq_sSFR}
\rm{log(sSFR)} = -0.65\,[\rm{log}(M_{\star})-10.31]-10.87 \quad,
\end{equation} 
where sSFR is in units of yr$^{-1}$ and the stellar mass is in units of M$_\odot$. The sSFR has been obtained from the MPA-JHU catalogue using a spectrophotometric synthesis fitting model. See \citet{2014ApJ...788...29L} for further details of these two relations. Following a similar criteria, galaxies above the relation are classified as star-forming (SF, 1209 SIG galaxies) and galaxies below as quenched (1087 SIG galaxies). There are three SIG galaxies with undefined sSFR, so we were not able to classify them.

Note that AGN classification based on purely emission-line BPT diagrams (as used in \citetalias{2013MNRAS.430..638S}) may be affected by uncertainties \citep{2012A&A...545A..15S}, where misclassified galaxies could be classified as LINERs or Seyferts. These uncertainties are specially important for massive main-sequence local galaxies that might be misclassified as passives \citep{2016MNRAS.457.2703R}. \citet{2016A&A...592A..30A} checked that less than the 4\% of the total number of galaxies in the SIG would be affected. 

At the same time, the selection of SFN galaxies is based on spectra obtained in the central fiber of the SDSS-DR7, which can be affected by AGN. We can obtain a more reliable selection of AGN and SFN galaxies combining the two methods (BPT diagrams and synthesis fitting model to the spectra). Therefore, we will study the fraction of AGN/SFN/passive galaxies as a function of the population of SF/quenched galaxies. Note also that a size-limited fiber leads to a different galaxy area coverage, thus fiber may cover some part of the disk for more distant galaxies. According to \citet{2003MNRAS.341...33K}, galaxies classified as SFN do not have more than a 1\% contribution of AGN, we are therefore confident that we are not affected by contamination of AGN in galaxies classified as SFN. However, we may misclassify galaxies with a star-forming disk as AGN. Given that galaxies with \citetalias{2013MNRAS.430..638S} classification are restricted to the local and narrow redshift range $0.03 \leq z \leq 0.08$, we do not expect a bias in our results caused by AGN misclassification or even possible redshift evolution.

As summary, the number of AGN, SFN, and passive galaxies in terms of morphology, colour and sSFR are collected in Table~\ref{tab:samples}. As expected, there is a small number of early-type, red, and quenched SFN SIG galaxies. Similarly, the number of late-type, blue, and star-forming isolated passives is also small.

\subsection{Quantification of the environment} \label{Sec:Qlss}

The public SIG catalogue also provides the isolation degree for SIG galaxies in terms of their large-scale environment \citep{2015A&A...578A.110A}. They use the tidal strength parameter \citep{2007A&A...472..121V,2013MNRAS.430..638S,2013A&A...560A...9A,2014A&A...564A..94A} to quantify the influence of the neighbour galaxies in their LSS, which are the galaxies lying within a volume of 5\,Mpc projected distance and line-of-sight velocity difference of $\Delta\,\varv~\leq~500$\,km\,s$^{-1}$. 

To study the influence of the large-scale environment on the AGN fraction for SIG galaxies, we therefore selected the $Q_{\rm LSS}$ parameter (Eq.~\ref{Eq:QLSS}), i. e. the tidal strength exerted by all the galaxies in the LSS. Then, for each galaxy, $i$, in the LSS at a projected distance $d_{LSS_i}$ and stellar mass $M_{LSS_i}$\footnote{Stellar masses for galaxies in the LSS were calculated by fitting the spectral energy distribution, on the five SDSS bands, using the routine kcorrect \citep{2007AJ....133..734B}.}, the total tidal strength on the SIG galaxy is:

\begin{equation} \label{Eq:QLSS}
Q_{\rm LSS} \equiv {\rm log} \left(\sum_i {\frac{M_{LSS_i}}{M}} \left(\frac{D}{d_{LSS_i}}\right)^3\right) \quad,
\end{equation} 
where $M$ and $D$\footnote{$D = 2\alpha r_{90}$ \citep{2013A&A...560A...9A}, where $r_{90}$, the Petrosian radius containing 90\,\% of the total flux of the galaxy in the $r$-band, is scaled by a factor $\alpha=1.43$ to recover the $D_{25}$, defined by the $\mu_B(B)$\,=\,25.0\,mag/arcsec$^2$ isophote.} are the stellar mass and the estimated diameter of the SIG galaxy, respectively. 

The greater the value of $Q_{\rm LSS}$, the less isolated from external influence the galaxy. In fact, using the visualization tool LSSGalPY \citep{2015ascl.soft05012A,2017PASP..129e8005A}, \citet{2016A&A...592A..30A} checked that SIG galaxies with low values of $Q_{\rm{LSS}}$ are generally located in voids ($Q_{\rm{LSS}}\,<\,-5.5$), and at higher values of the $Q_{\rm{LSS}}$ the galaxies are more related with denser structures, e. g., in the outskirts of clusters ($Q_{\rm{LSS}}\,>\,-4.5$). In this regard, it is appropriate to note SIG galaxies mainly belong to the outer parts of filaments, walls, and clusters, and generally differ from the void population of galaxies, where only one-third of SIG galaxies are located in voids \citep{2015A&A...578A.110A}.

\section{Results} \label{Sec:res}

\subsection{AGN prevalence in isolated galaxies} \label{Sec:res1}

We compare the relative fraction of SFN, AGN, and passive galaxies with respect to the stellar mass in the SIG, as a function of morphology, colour, and sSFR. Since there is a strong dependence of AGN with stellar mass, it is recommendable to made a separated study in different stellar mass bins \citepalias{2013MNRAS.430..638S}. According to the distributions shown in Fig.~\ref{Fig:mass}, we separate into seven stellar mass bins from $\rm{log}(M_\star)~=~10.0~[M_\odot]$ to $\rm{log}(M_\star)~=~11.4~[M_\odot]$.

\begin{figure*}
\centering
\includegraphics[width=\textwidth]{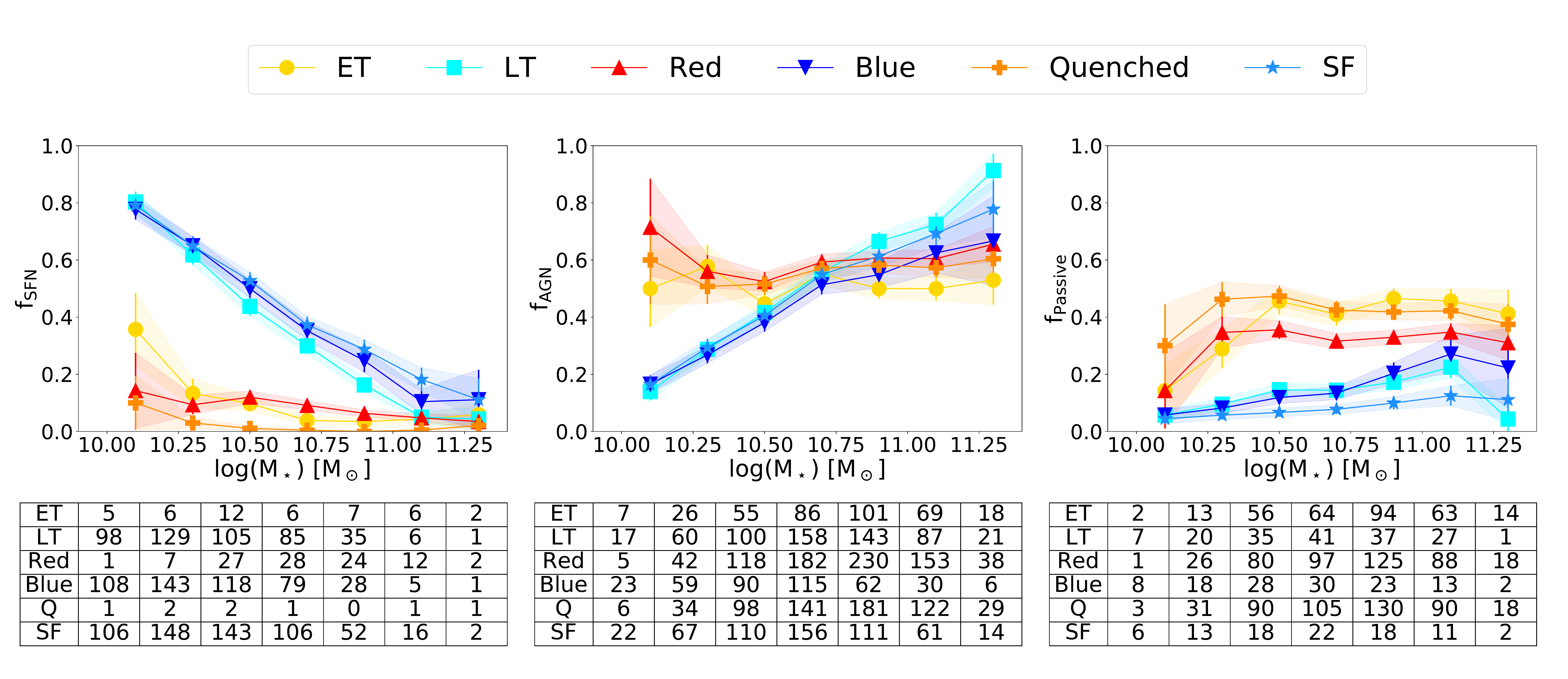}
\caption{Fraction of SFN (left panel), optical AGN (middle panel), and passive galaxies (right panel) with respect to stellar mass. The fraction in early-type ($N_T~=~712$), red ($N_T~=~1304$), and quenched ($N_T$~=~1087) SIG galaxies is depicted by yellow circles, red triangles, and orange pluses, respectively. Cyan squares correspond to the fraction in late-type SIG galaxies ($N_T$~=~1218), blue inverted triangles for blue SIG galaxies ($N_T$~=~995), and light blue stars for star-forming SIG galaxies ($N_T$~=~1209). The number of galaxies in each stellar mass bin is shown in tables for each sample at the bottom of each panel. Error bars are given by considering binomial distribution.}
\label{Fig:massbins}
\end{figure*}

The left panel in Fig.~\ref{Fig:massbins} shows that the fraction of SFN decreases with stellar mass for late-type, blue, and star-forming SIG galaxies, where the fraction for late-type galaxies is slightly steeper. On the contrary, the fraction of AGN increases with stellar mass (middle panel in Fig.~\ref{Fig:massbins}) for these galaxies. The fraction of passives also increases but the tendency is less steep, where the fraction for star-forming isolated galaxies is the smoothest one (right panel in Fig.~\ref{Fig:massbins}). The same trend of AGN with stellar mass is observed if we separate late-type, blue, and star-forming SIG galaxies in TOs, Seyferts and LINERs (see Fig.~\ref{Fig:AppFig2AGN}).

The fractions of SFN, AGN, and passives for early-type, red, and quenched SIG galaxies are roughly constant with stellar mass, with except to the first bin, where the number of these galaxies at low stellar mass is small.

As shown in the left panel in Fig.~\ref{Fig:massbins}, in general the fraction of late-type, blue, and star-forming SIG galaxies classified as SFN in each stellar mass bin is larger than the fraction of early-type, red, quenched SIG galaxies, but is smaller for galaxies classified as passives as shown in the right panel. In the case of galaxies classified as AGN, the fraction of late-type, blue, and star-forming galaxies is in general higher or equal than that of early-type, red, and quenched SIG galaxies at stellar masses $\rm{log}(M_\star)~>~10.6~[M_\odot]$.

Note that the observed trends may be affected by the low number of early-type, red, and quenched galaxies at low stellar mass, and late-type, blue and star-forming galaxies at high stellar mass (see also the stellar mass distributions in Fig.~\ref{Fig:mass}). For that reason we focus only on general trends observed in bins with a larger statistic. 

\subsection{Dependence on the LSS environment} \label{Sec:res2}

As introduced in Sect.~\ref{Sec:Qlss}, we use the $Q_{\rm LSS}$ to quantify the effect of the LSS on the fractions of SFN, AGN, and passive galaxies in the SIG. To take into account the effect of the mass and having enough statistic, we divide the samples into two stellar mass bins: low-mass galaxies ($10.0~\leq~\rm{log}(M_\star)~<~10.6~[M_\odot]$) and high-mass galaxies ($10.6~\leq~\rm{log}(M_\star)~\leq~11.4~[M_\odot]$). We selected this mass cut value of $\rm{log}(M_\star)~=~10.6~[M_\odot]$ to have a similar number of galaxies at both sides of the distributions (see Fig.~\ref{Fig:mass}) and also it is approximately the mass value where the fraction of AGN for early-type, red, and quenched SIG galaxies matches the fraction of AGN for late-type, blue, star-forming SIG galaxies.  

\begin{figure*}
\centering
\includegraphics[width=\textwidth]{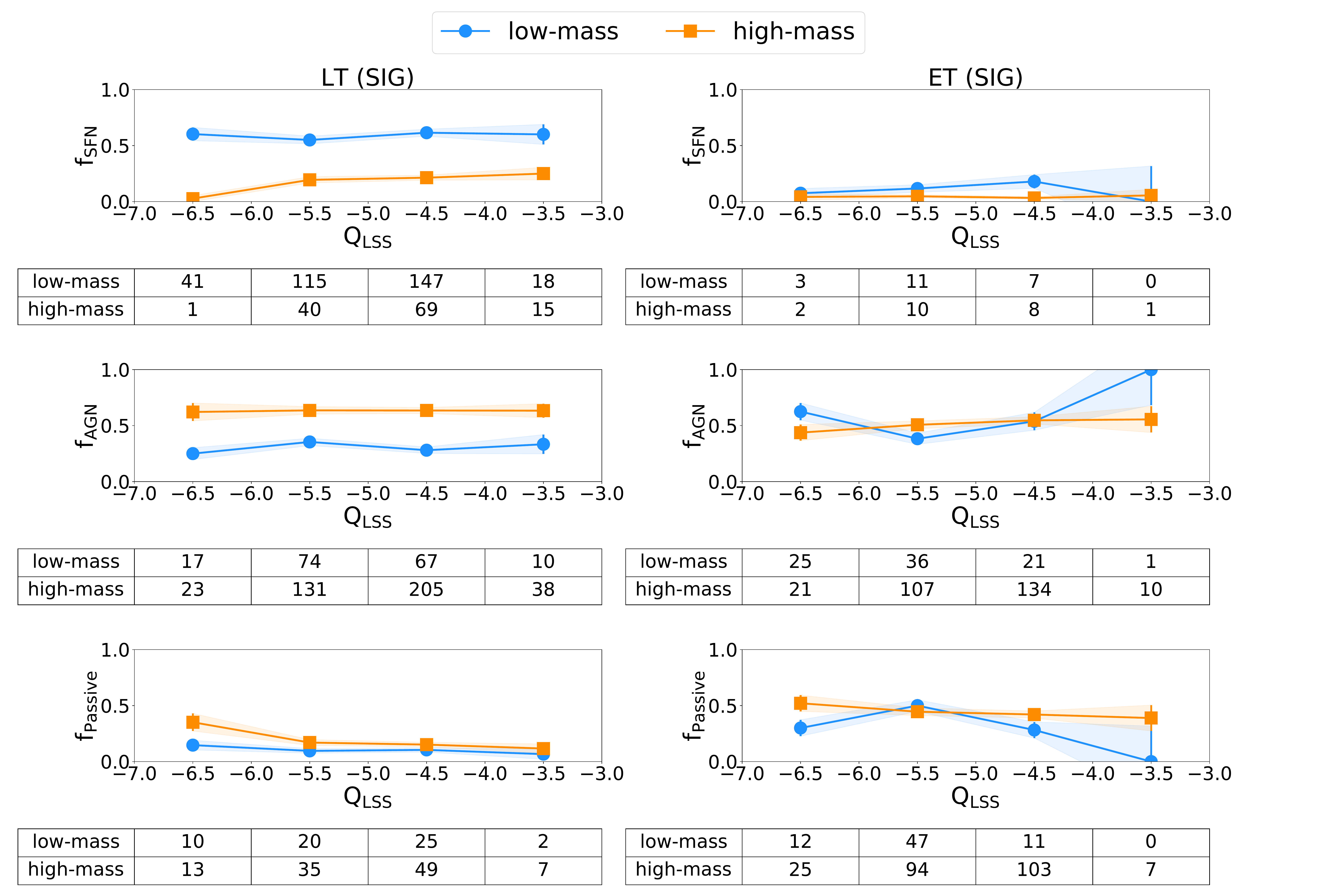}
\caption{Fraction of optical nuclear activity with respect to the $Q_{\rm LSS}$ environmental parameter according to their morphology. Low-mass galaxies ($10.0~\leq~\rm{log}(M_\star)~<~10.6~[M_\odot]$) are represented by cyan circles, and high-mass galaxies ($10.6~\leq~\rm{log}(M_\star)~\leq~11.4~[M_\odot]$) are represented by orange squares. The fraction of SFN, AGN, and passive late-type SIG galaxies ($N_T$ = 462, 586, and 170, respectively) is represented from top to bottom in the left panels, and for early-type SIG galaxies ($N_T$ = 44, 362, and 306, respectively) in the right panels. The number of galaxies in each $Q_{\rm LSS}$ bin is shown in tables for each sample at the bottom of each panel. Error bars are given considering binomial distribution.}
\label{Fig:SIG_LSS_morpho_106}
\end{figure*}

\begin{figure*}
\centering
\includegraphics[width=\textwidth]{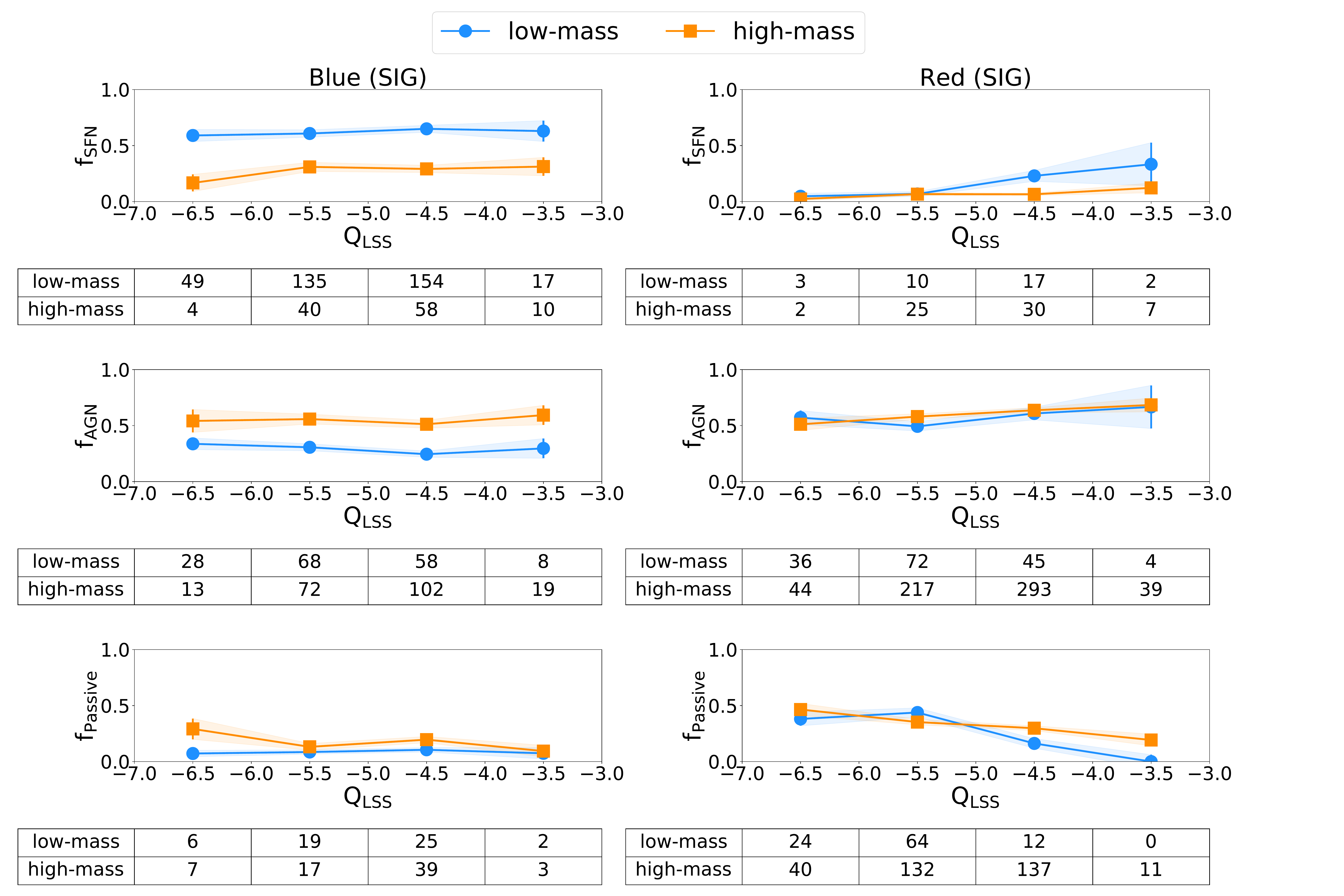}
\caption{Fraction of optical nuclear activity with respect to the $Q_{\rm LSS}$ environmental parameter according to their colour. Low-mass galaxies ($10.0~\leq~\rm{log}(M_\star)~<~10.6~[M_\odot]$) are represented by cyan circles, and high-mass galaxies ($10.6~\leq~\rm{log}(M_\star)~\leq~11.4~[M_\odot]$) are represented by orange squares. The fraction of SFN, optical AGN, and passive blue SIG galaxies ($N_T$ = 486, 385, and 124, respectively) is represented from top to bottom in the left panels, and for red SIG galaxies ($N_T$ = 101, 768, and 435, respectively) in the right panels. The number of galaxies in each $Q_{\rm LSS}$ bin is shown in tables for each sample at the bottom of each panel. Error bars are given considering binomial distribution.}
\label{Fig:SIG_LSS_color_106}
\end{figure*}

\begin{figure*}
\centering
\includegraphics[width=\textwidth]{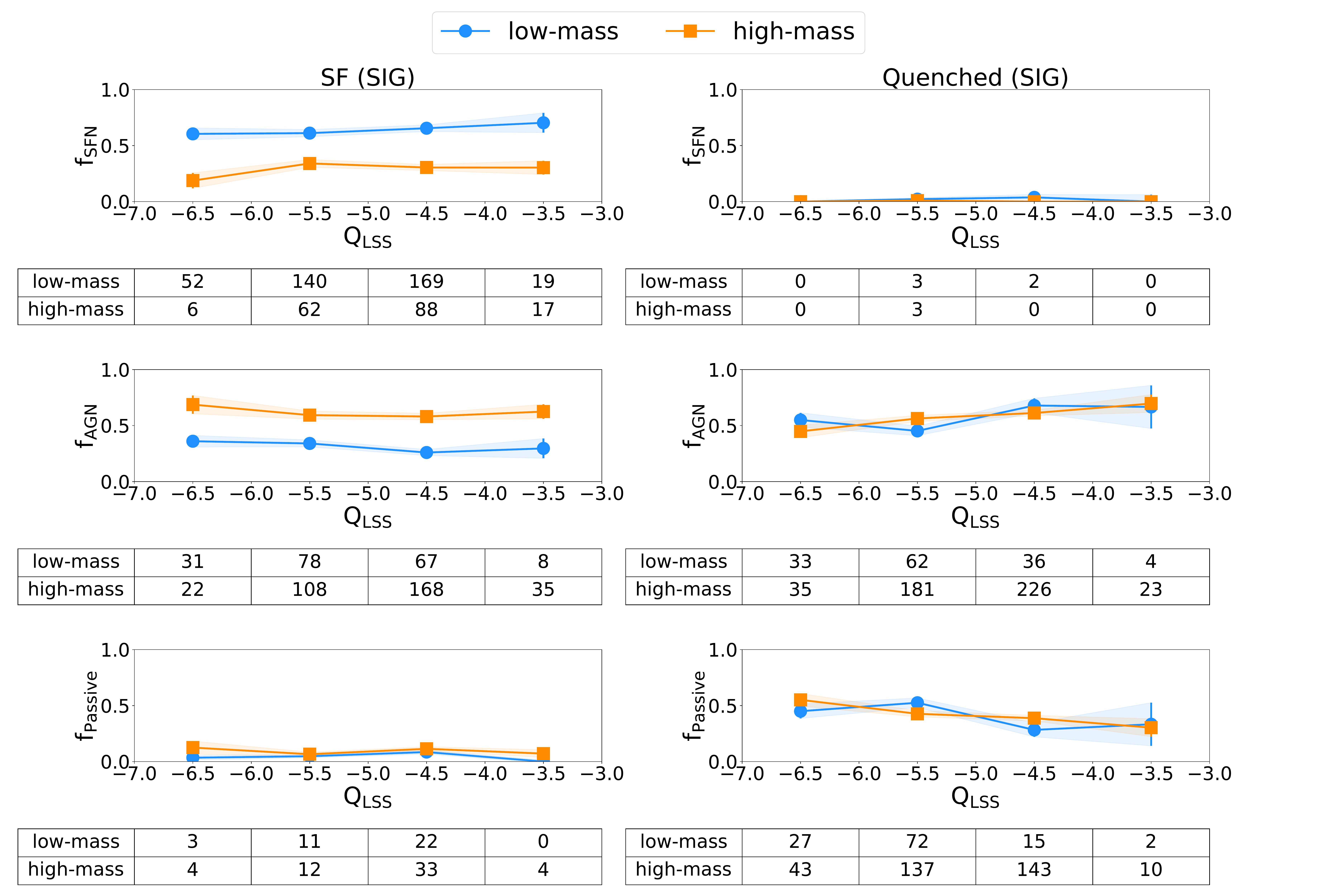}
\caption{Fraction of optical nuclear activity with respect to the $Q_{\rm LSS}$ environmental parameter according to their sSFR status. Low-mass galaxies ($10.0~\leq~\rm{log}(M_\star)~<~10.6~[M_\odot]$) are represented by cyan circles, and high-mass galaxies ($10.6~\leq~\rm{log}(M_\star)~\leq~11.4~[M_\odot]$) are represented by orange squares. The fraction of SFN, optical AGN, and passive star forming SIG galaxies ($N_T$ = 577, 541, and 91, respectively) is represented from top to bottom in the left panels, and for quenched SIG galaxies ($N_T$ = 8, 611, and 468, respectively) in the right panels. The number of galaxies in each $Q_{\rm LSS}$ bin is shown in tables for each sample at the bottom of each panel. Error bars are given considering binomial distribution.}
\label{Fig:SIG_LSS_status_106}
\end{figure*}

Figure~\ref{Fig:SIG_LSS_morpho_106} shows the fraction of optical nuclear activity, segregated in low- and high-mass bins, with respect to $Q_{\rm LSS}$, for SIG galaxies as a function of their morphology. Higher values of the $Q_{\rm LSS}$ correspond to a stronger interaction with the LSS. Similarly, figures~\ref{Fig:SIG_LSS_color_106} and \ref{Fig:SIG_LSS_status_106} show the fraction of SFN, AGN, and passive SIG galaxies in terms of colour and sSFR, respectively.

In general, there is no clear dependence on the large-scale environment in the nuclear activity for SIG galaxies in terms of morphology and colour, although there is a trend for high-mass SIG galaxies to increase with denser LSS environment. There are some trends in terms of sSFR classification, which will be discussed in Sect.~\ref{Sec:dis}, as well as other general trends.

\section{Discussion}  \label{Sec:dis}

\subsection{AGN prevalence in isolated galaxies} \label{Sec:dis1}

It is known that there is a strong dependence of the prevalence of AGN on the mass of the host galaxy \citep{2003MNRAS.346.1055K,2004MNRAS.353..713K,2005MNRAS.362...25B}. Following the previous work in \citet{2016A&A...592A..30A}, we control stellar mass restricting the galaxies to the stellar mass range $10.0~\leq~\rm{log}(M_\star)~\leq~11.4~[M_\odot]$. They found that, for any of the galaxy samples considered in their study, the fraction of SFN decreases with stellar mass, and the fractions of AGN and passives increase with higher stellar mass (where the trend is steeper for AGNs). In the present work we explore these trends in isolated galaxies as a function of their morphology, colour, and sSFR. 

As it is shown in Fig.~\ref{Fig:massbins}, the previous trends are only observed in late-type, blue, and star-forming isolated galaxies. However, the fractions for early-type, red, and quenched isolated galaxies are roughly constant and independent of the stellar mass. This means that once an isolated galaxy is in a quiescent state (evolved, without forming new stars), there is about 50\% to 60\% probability that the galaxy hosts an AGN and about 40\% probability that is a passive galaxy, independently of its stellar mass. \citet{2016A&A...592A..30A} did not observe this bimodality since most of the SIG (about 75\%) is composed of late-type galaxies, in agreement to the morphology-density relation.

According to the central panel in Fig.~\ref{Fig:massbins}, there is a higher fraction of AGNs in low-mass early-type, red, and quenched isolated galaxies than for low-mass late-type, blue, and star-forming isolated galaxies, in agreement with \citet{2012A&A...545A..15S} and \citet{2016MNRAS.459..291H}, which suggest that the role of a bulge, and a large gas reservoir are both essential to trigger optical nuclear activity. Once the galaxy reaches a stellar mass about $\rm{log}(M_\star)~\sim~10.6~[M_\odot]$, the probability that an isolated galaxy hosts an AGN is independent of its morphology, colour, or sSFR. At higher stellar mass, the fraction of AGN in late-type, blue, and star-forming isolated galaxies is larger. This may be affected the fact that galaxies with stellar masses $\rm{log}(M_\star)~>~10.5~[M_\odot]$ may present large bulges, therefore they show lower star formation and favour AGN classification in a BPT diagram \citep{2004MNRAS.351.1151B,2016MNRAS.455.2826R}. In fact \citet{2017A&A...599A..71D} observe a decrease of the SFR in galaxies with $\rm{log}(M_\star)~>~10.6~[M_\odot]$. 

For consistency with \citet{2016A&A...592A..30A}, and to have a statistically meaningful number of galaxies, we consider TO, Seyfert, and LINER as AGN galaxies following also the work of \citet{2013MNRAS.430..638S}. However, recent studies of the spatially resolved nebular emission in nearby galaxies, using Integral Field Unit (IFU) data, suggest that LINER-like emission is more related to evolved stars and not directly related to nuclear ionization \citep[e.g.,][]{2013A&A...558A..43S,2016MNRAS.461.3111B}. In this regard, we explore the effect of considering LINER galaxies the AGN classification.

\begin{figure}
\centering
\includegraphics[width=\columnwidth]{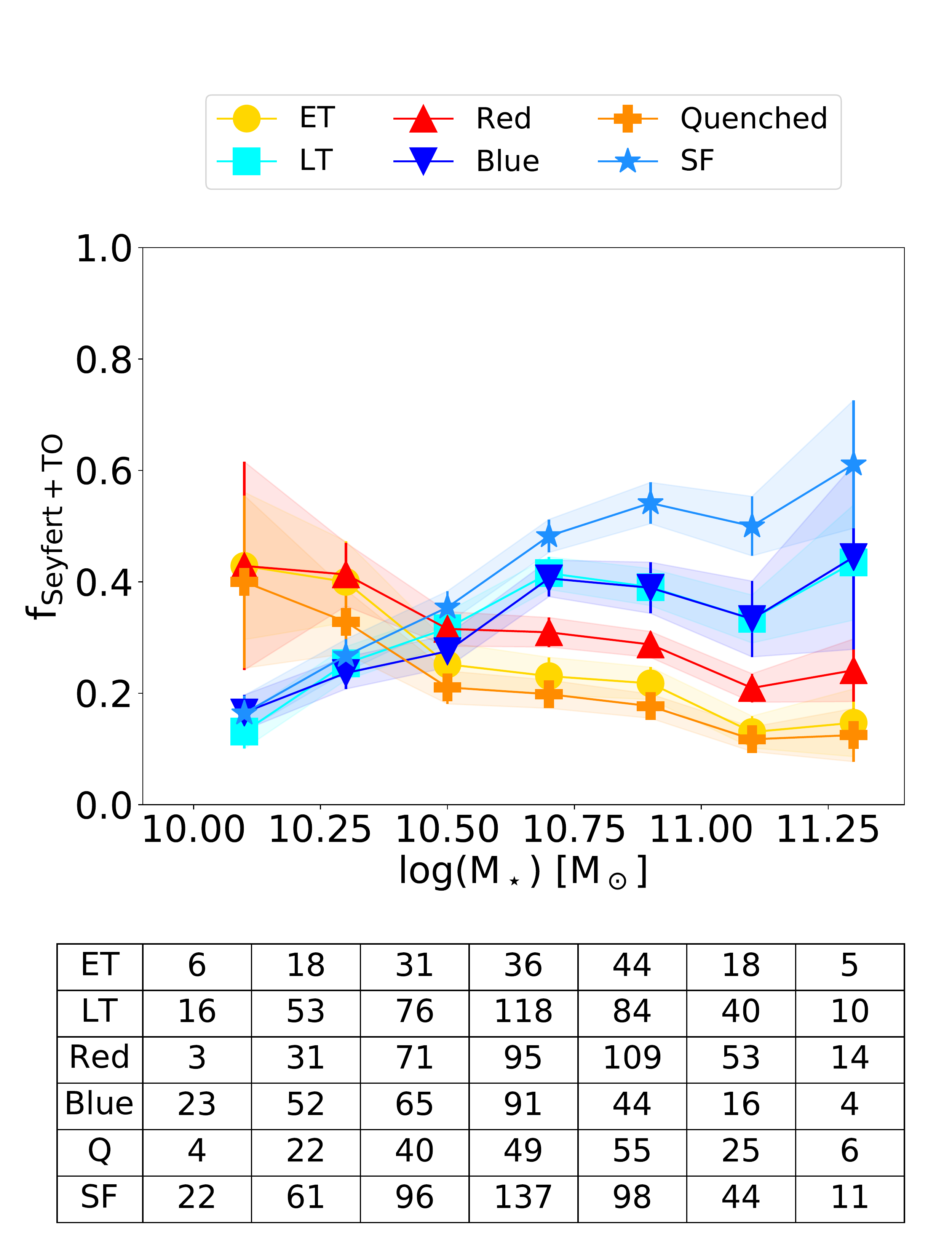}
\caption{Similar to the middle panel in the Fig.~\ref{Fig:massbins} but considering only TO and Seyfert as AGN subtypes (i.e., excluding SIG galaxies classified as LINER). The number of galaxies in each stellar mass bin is shown in the table at the bottom. Error bars are given by considering binomial distribution.}
\label{Fig:massbins4}
\end{figure}

Similar to the central panel in the Fig.~\ref{Fig:massbins}, the Fig.~\ref{Fig:massbins4} shows the fraction of AGN when considering only Seyfert and TO galaxies. We observe similar trends under this consideration, i.e. the fraction of AGN  increases with stellar mass in late-type, blue and star-forming SIG galaxies, which is higher than the fraction in early-type, red, and Quenched SIG galaxies at higher masses. However we observe that this happens at slightly smaller stellar mass ($\rm{log}(M_\star)~>~10.5~[M_\odot]$). This value was also considered to separate between low- and high-mass galaxies when exploring the effect of the LSS environment in the fraction of AGN (see Appendix~\ref{Sec:App}), and it is therefore considered in the range of uncertainties. The most important difference when excluding LINER galaxies is the slow decrement of the fraction of AGN in early-type, red, and quenched galaxies with stellar mass, instead of being roughly constant.

Note that we do not exclude TOs in the AGN classification, since they contain AGN component up to 40\% of their H$\alpha$ luminosity \citep{2004MNRAS.351.1151B}. Given that our analysis is based on spectroscopic data from the 3'' fibre at the central area of the galaxies, which typically contains the nuclear region at the redshift range considered in our galaxy sample, we can consider that the effect of non-AGN emission in our classification is minimal. However, we additionally explore the AGN prevalence in each AGN subtype i.e., TOs, Seyferts, and LINERs. The results of this analysis are presented in the Appendix~\ref{Sec:AppB}. Even if the statistic is low, the general trend for late-type, blue, and star-forming galaxies is to increase with the stellar mass (see Fig.~\ref{Fig:AppFig2AGN}), in agreement with the results shown in the middle panel of Fig.~\ref{Fig:massbins}. The TO and Seyfert prevalence for early-type, red, and quenched SIG galaxies is roughly constant, also in agreement with the previous results. Only when considering LINERs we observe an increment with stellar mass. However, the fractions of LINERs at low mass for early-type, red and quenched galaxies are significantly larger than late-type, blue, and star-forming galaxies, in agreement with the previous results. Therefore, our main conclusions overall are valid.

\subsection{Dependence on the LSS environment}  \label{Sec:dis2}

\citet{2016A&A...592A..30A} found a strong dependence of the LSS on the optical nuclear activity and star formation in isolated galaxies, where the observed trends are different depending on galaxy mass. In particular, they found an increment of the fraction of low-mass SFN isolated galaxies with denser LSS. This trend is reproduced by low-mass star-forming isolated galaxies (see the upper left panel in Fig.~\ref{Fig:SIG_LSS_status_106}), which is even steeper if we select a mass cut of $\rm{log}(M_\star)~=~10.5~[M_\odot]$ (see the upper left panel in Fig.~\ref{Fig:App105}) instead of the selected $\rm{log}(M_\star)~=~10.6~[M_\odot]$. Surprisingly, we also observe this trend for low-mass red isolated galaxies (see the upper right panel in Fig.~\ref{Fig:SIG_LSS_color_106}), even if there is a low fraction of red SFN galaxies. 

To explore the connection on this trend for these two different population of galaxies, we have checked that low-mass red isolated galaxies with a SFN at high $Q_{\rm LSS}$ are also star-forming galaxies. This population of red and star-forming galaxies in the Local Universe is known as Red Misfits \citep{2018MNRAS.476.5284E}, a transition population to the red sequence, where the quenching is dominated by internal processes rather than environmentally-driven processes. Then it would be somehow expected to find this population in isolated galaxies. However \citet{2018MNRAS.476.5284E} found that the proportion of Red Misfits is nearly independent of environment, and we find this population for isolated galaxies at denser LSS. These preliminary differences are worth a further detailed study.

We do not see any trend in SFN for low-mass isolated galaxies in terms of morphology (Fig.~\ref{Fig:SIG_LSS_morpho_106}). In the case of high-mass isolated galaxies, in general, there is no dependence of the LSS on the fraction of SFN, although it seems there is a slight tendency for late-type isolated galaxies to increases with the $Q_{\rm LSS}$ (see the upper left panel in Fig.~\ref{Fig:SIG_LSS_morpho_106}). 

As expected, the fraction of SFN in late-type, blue, and star-forming galaxies is generally larger than in early-type, red, and quenched galaxies, especially in low-mass galaxies (see also the left panel in Fig.~\ref{Fig:massbins}). With except to low-mass star-forming galaxies, where the fraction increases with higher $Q_{\rm LSS}$, the fractions of SFN are independent of the LSS. Therefore, even when exploring the trends observed in \citet{2016A&A...592A..30A} in terms of morphology, colour, and sSFR, we still do not observe an increment of the SFN in void galaxies (here SIG galaxies with low values of the $Q_{\rm LSS}$), as reported in \citet{2015ApJ...810..165L}. The difference between the results of the two studies may be therefore caused by the fact that the void population of galaxies are composed of galaxies with diverse local environments, commonly known as field galaxies, which may be composed of pairs and small groups of galaxies, where the spectral properties may be different and be affected by the group environment and/or interactions. 

\citet{2016A&A...592A..30A} also found an increment of the fraction of AGN in high-mass SIG galaxies, and a decrement in low-mass galaxies, with denser LSS. Our results reproduce these trends in terms of the colour and the sSFR, in particular for high-mass red and quenched isolated galaxies, and low-mass blue and star-forming isolated galaxies, respectively (see the middle panels in Figures~\ref{Fig:SIG_LSS_color_106} and \ref{Fig:SIG_LSS_status_106}). Moreover, the increment of the fraction of AGN with denser LSS seems to be independent of the stellar mass for both red and quenched galaxies. In addition, we also observe the fraction of AGN in high-mass early-type isolated galaxies slightly increases with $Q_{\rm LSS}$ (see the right middle panel in Fig.~\ref{Fig:SIG_LSS_morpho_106}). 
On the contrary, the fraction of AGN in high-mass late-type, blue, and star-forming isolated galaxies does not depend on the LSS environment. So we can conclude that the AGN fractions for late-type, blue, and star-forming isolated galaxies depend strongly on the stellar mass, i.e. AGN is ``mass triggered''.

Regarding passive galaxies, the results are in agreement with \citet{2016A&A...592A..30A}, where the general trend of the fraction of passive isolated galaxies is to decrease with higher value of the $Q_{\rm LSS}$, i.e. from voids to denser regions as clusters and filaments. This trend is also independent of galaxy mass, morphology, colour, and sSFR. 

As introduced in Sec.~\ref{Sec:AGN}, the classification of AGN based on BPT diagrams is not perfect and there might be some issues to take into consideration. The obtained results, in comparison to \citet{2016A&A...592A..30A}, show that a separation between quenched and star-forming galaxies in terms of the sSFR provides a clean selection of SFN and AGN galaxy populations, since some star-forming galaxies could be classified as AGNs using BPT diagrams. Note also the sensitivity of the stellar mass cut $\rm{log}(M_\star)~=~10.6~[M_\odot]$ on the fraction of AGN in terms of sSFR when separating between low-mass and high-mass galaxies (see Appendix \ref{Sec:App} for results using other stellar mass cuts). On the other hand, the observed trends in terms of morphology and colour are robust. The ``mass triggered'' AGN are still dominant in late-type, blue, and star-forming SIG galaxies for different stellar mass cuts.

We also found that, in terms of colour, isolated galaxies are sensitive to the LSS, where we can also confirm that the fraction of AGN increases with denser LSS. This is more evident in red galaxies, where this increment is also independent of the stellar-mass. A similar dependence on the environment in terms of colour was reported by \citet{2011MNRAS.417..453D} for star formation rate and sSFR, where the dependence is stronger for red galaxies. In the present work we report for the first time the environmental dependence on AGN in terms of colour, for isolated galaxies with respect to their large-scale environment, independently of their stellar mass. The results suggest that the secular processes that trigger AGN on red and quenched isolated galaxies, and partially for early-type galaxies where the trends are a bit more noisy, are specially sensitive to the LSS, i.e. ``environment triggered'', increasing their fraction and probably being feeded by the accretion of cold gas from the LSS. 

\section{Summary and conclusions} \label{Sec:con}

Galaxy mass and environment are both playing a major role in the evolution of galaxies \citep{2010ApJ...721..193P,2012ApJ...757....4P}. At the same time, AGN has been proposed as a plausible physical mechanism of the evolution of star-forming galaxies in quenched galaxies. Accordingly, we try to identify the connections between AGN activity and the stellar mass and the large-scale environment (LSS) in a population of isolated galaxies, where AGN would not be triggered by recent galaxy interactions or mergers, and thus black hole growths by secular processes \citep{2013MNRAS.433.1479H}. In this work we extend a previous study carried out by \citet{2016A&A...592A..30A} and we focus on isolated galaxies to investigate the effect of both, mass and environment, on the fraction of optical nuclear activity in isolated galaxies as a function of their morphology, colour, and sSFR. 

We selected galaxies in the the SDSS-based catalogue of isolated galaxies \citep[SIG,][]{2015A&A...578A.110A}. We used the AGN classification provided by \citep{2013MNRAS.430..638S}, based on BPT diagnosis diagrams, where AGN are composed of TO, Seyfert, and LINER galaxies. We divided our galaxy sample in terms of morphology \citep[provided by][]{2011A&A...525A.157H}, colour, and sSFR \citep[according to the relations for SDSS galaxies provided by][]{2014ApJ...788...29L}. To quantify the effect of the LSS around SIG galaxies, we used the tidal strength exerted by all the neighbour galaxies in a volume of 5\,Mpc field radius within 500\,km\,s$^{-1}$ line-of-sight velocity difference \citep{2014A&A...564A..94A,2015A&A...578A.110A}. 

In Sect.~\ref{Sec:res1} we explored the connection between AGN and stellar mass when separating isolated galaxies in terms of morphology, colour and sSFR. In Sect.~\ref{Sec:res2} we fixed the stellar mass into low- and high-mass galaxies ($10.0~\leq~\rm{log}(M_\star)~<~10.6~[M_\odot]$ and $10.6~\leq~\rm{log}(M_\star)~\leq~11.4~[M_\odot]$, respectively) to explore where AGN activity is affected by the LSS. 

Our findings are the following:

\begin{enumerate}
\item We found that AGN is strongly affected by stellar mass in  ''active'' galaxies (namely late-type, blue, and star-forming), however it has no influence for ``quiescent'' galaxies (namely early-type, red, and quenched).

\item In agreement with \citet{2016A&A...592A..30A}, the fraction of SFN decreases with stellar mass and the fraction of AGN increases, but in the present study we only observe this trend in ``active'' isolated galaxies. 

\item On the contrary, when an isolated galaxy is in a quiescent state, the fraction of SFN, AGN, and passive isolated galaxies is independent of the stellar mass. 

\item The fraction of AGN in low-mass ''quiescent'' isolated galaxies is higher than that in late-type, blue, or star-forming isolated galaxies of the same mass. Once the galaxy reaches a stellar mass about $\rm{log}(M_\star)~\sim~10.6~[M_\odot]$, the probability that an isolated galaxy hosts an AGN is independent of its morphology, colour, or sSFR. 

\item In general, the trends previously found in \citet{2016A&A...592A..30A} with respect to the LSS are weakly reproduced for ''active'' isolated galaxies but well reproduced by ``quiescent'' galaxies in terms of sSFR and colour. We do not find a clear dependency on the LSS in the fraction of nuclear activity with galaxy morphology. 

\item In comparison to \citet{2016A&A...592A..30A}, we find an increment on the fraction of SFN with denser LSS in low-mass star forming and red isolated galaxies. We find that these low-mass red galaxies with a SFN at high $Q_{\rm LSS}$ are also star-forming galaxies, which would be considered in the population of Red Misfits \citep{2018MNRAS.476.5284E}. 

\item Regarding AGN, we find a clear increment of the fraction of AGN with denser environment in quenched and red isolated galaxies, independently of the stellar mass. However we do not find a decrement in low-mass red and quenched galaxies, but this is observed in low-mass blue and star-forming SIG galaxies.
\end{enumerate}

This clear separability of the effects of environment and stellar mass on nuclear activity suggests that there are two distinct processes at work. Thereby AGN activity would be ``mass triggered'' in ``active'' isolated galaxies. This means that AGN is independent of the intrinsic property of the galaxies, except on its stellar mass. On the other hand, AGN would be ``environment triggered'' in ``quiescent'' isolated galaxies, where the fraction of AGN in terms of sSFR and colour increases from void regions to denser LSS, independently of its stellar mass. 

\section*{Acknowledgments}

The authors acknowledge the anonymous referee for his/her report, which helped to clarify and improve the quality of this work.

This work was supported by CONICYT Astronomy Program CAS-CONICYT project No.\,CAS17002. This work is sponsored by the Chinese Academy of Sciences (CAS), through a grant to the CAS South America Center for Astronomy (CASSACA) in Santiago, Chile. MAF is grateful for financial support from CONICYT FONDECYT project No.\,3160304. IL acknowledges partial financial support from PROYECTO FONDECYT REGULAR 1150345. SDP acknowledge financial support from the Spanish Ministerio de Economía y Competitividad under grant AYA2013-47742-C4-1-P, AYA2017-79724-C4-4-P from the Spanish PNAYA, and from Junta de Andaluc\'ia Excellence Project PEX2011-FQM-7058.

This research made use of \textsc{astropy}, a community-developed core \textsc{python} ({\tt http://www.python.org}) package for Astronomy \citep{2013A&A...558A..33A}; \textsc{ipython} \citep{PER-GRA:2007}; \textsc{matplotlib} \citep{Hunter:2007}; \textsc{numpy} \citep{:/content/aip/journal/cise/13/2/10.1109/MCSE.2011.37}; \textsc{scipy} \citep{citescipy}; and \textsc{topcat} \citep{2005ASPC..347...29T}.

Funding for SDSS-III has been provided by the Alfred P. Sloan Foundation, the Participating Institutions, the National Science Foundation, and the U.S. Department of Energy Office of Science. The SDSS-III web site is http://www.sdss3.org/. 
SDSS-III is managed by the Astrophysical Research Consortium for the Participating Institutions of the SDSS-III Collaboration including the University of Arizona, the Brazilian Participation Group, Brookhaven National Laboratory, University of Cambridge, University of Florida, the French Participation Group, the German Participation Group, the Instituto de Astrofisica de Canarias, the Michigan State/Notre Dame/JINA Participation Group, Johns Hopkins University, Lawrence Berkeley National Laboratory, Max Planck Institute for Astrophysics, New Mexico State University, New York University, Ohio State University, Pennsylvania State University, University of Portsmouth, Princeton University, the Spanish Participation Group, University of Tokyo, University of Utah, Vanderbilt University, University of Virginia, University of Washington, and Yale University.

\bibliographystyle{aa}
\bibliography{references}

\begin{appendix} 

\section{AGN subtypes prevalence}
\label{Sec:AppB}

The AGN classification used in this study, adopted from \citet{2013MNRAS.430..638S}, is composed of galaxies classified as TO, Seyfert, and LINER. In Sec.~\ref{Sec:res1} we studied the optical AGN prevalence of isolated galaxies with respect to the stellar mass, as a function of morphology, colour, and sSFR. To explore the effect of the AGN classification in our conclusions, in this section we compare the relative fraction of each AGN subtype at the same stellar mass bins.

\begin{figure*} 
\centering
\includegraphics[width=\textwidth]{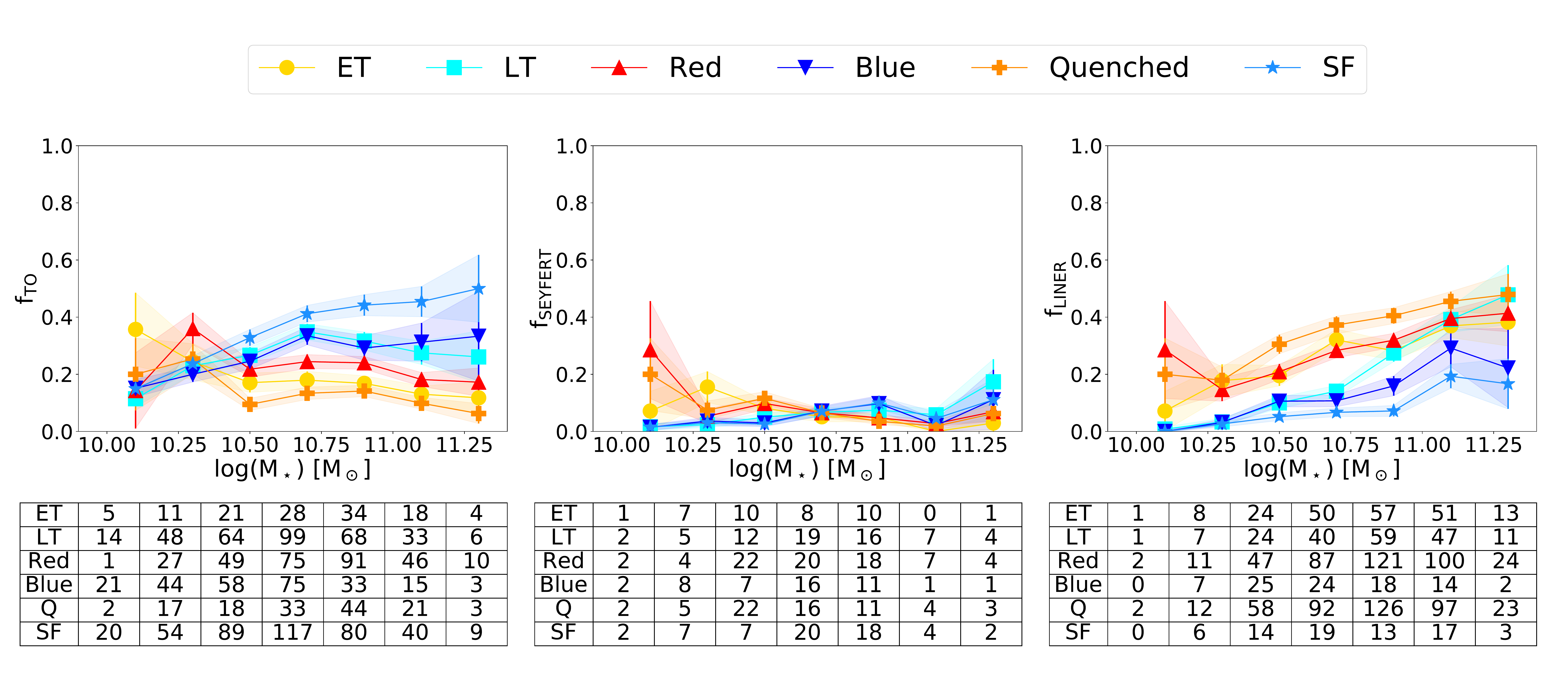}
\caption{
Fraction of TO (left panel), Seyfert (middle panel), and LINER galaxies (right panel) with respect to stellar mass. Following the Fig.~\ref{Fig:massbins}, the fraction in early-type ($N_T~=~362$), red ($N_T~=~768$), and quenched ($N_T$~=~611) SIG galaxies classified as AGN are depicted by yellow circles, red triangles, and orange pluses, respectively. Cyan squares correspond to the fraction of AGN in late-type SIG galaxies ($N_T$~=~586), blue inverted triangles for blue SIG galaxies ($N_T$~=~385), and light blue stars for star-forming SIG galaxies ($N_T$~=~541). The fractions of TO (left panel), Seyfert (middle panel), and LINER (right panel) galaxies are shown in tables for each sample at the bottom of each panel. Error bars are given by considering binomial distribution.
}
\label{Fig:AppFig2AGN}
\end{figure*}

The three panels of Fig.~\ref{Fig:AppFig2AGN} show the fractions of SIG galaxies classified as TO, Seyfert, and LINER. Therefore, at fixed stellar mass, the sum of each number in the lower tables is equal to the number of AGN in the lower table of the middle panel in Fig.~\ref{Fig:massbins} for the same galaxy group. For instance, the sum of the number of isolated early-type galaxies classified as TO (5 galaxies), Seyfert (1 galaxy), and LINER (1 galaxy) in the first mass bin is equal to 7, as shown in the table of the middle panel in Fig.~\ref{Fig:massbins}.

The trend observed in the AGN fractions of massive galaxies, $\rm{log}(M_\star)~>~10.6~[M_\odot]$, in the middle panel of Fig.~\ref{Fig:massbins} is qualitatively reproduced by ``active'' (late-type, blue, and star-forming) SIG galaxies with equal or larger fractions of TOs and Seyferts than ``passive'' (early-type, red, and quenched) SIG galaxies. In the case of LINERs, the fraction is slightly lower in active galaxies than in passive galaxies. For lower stellar masses, the middle panel of Fig.~\ref{Fig:massbins} shows an overall trend where the fraction of each AGN subtype in passive galaxies are equal or larger than those of active galaxies.  This trend, even if is noisier, is still observable when we separate the AGN classification in the three subtypes.

The possible consequences of these results on our main conclusions are discussed in Sec.~\ref{Sec:dis1}.

\section{Effect of the selected stellar mass cut}
\label{Sec:App}

According to the distributions presented in Fig.~\ref{Fig:mass}, we selected a stellar mass cut $\rm{log}(M_\star)~=~10.6~[M_\odot]$ as a good compromise between the number of galaxies at each side of the distributions and an equally valid value when separating between late- and early-type galaxies, blue and red galaxies, and star-forming and quenched galaxies. 

However, when exploring the AGN prevalence as a function of the stellar mass, we observe that there is a transition stellar mass range around the selected mass cut $\rm{log}(M_\star)~=~10.6~[M_\odot]$, where the fractions of AGN in ''active'' isolated galaxies equal the fractions of AGN in ''quiescent'' isolated galaxies (see the middle panel in Fig.~\ref{Fig:massbins}).

Here we explore whether there is any effect of the stellar mass cut value on the observed trends of the fractions of SFN, AGN, and passive galaxies with the LSS, presented in Sect.~\ref{Sec:dis2}. We select $\rm{log}(M_\star)~=~10.5~[M_\odot]$ and $\rm{log}(M_\star)~=~10.7~[M_\odot]$ as a low and high limit values of the critical stellar mass region. 

\begin{figure*}
\centering
\includegraphics[width=\textwidth]{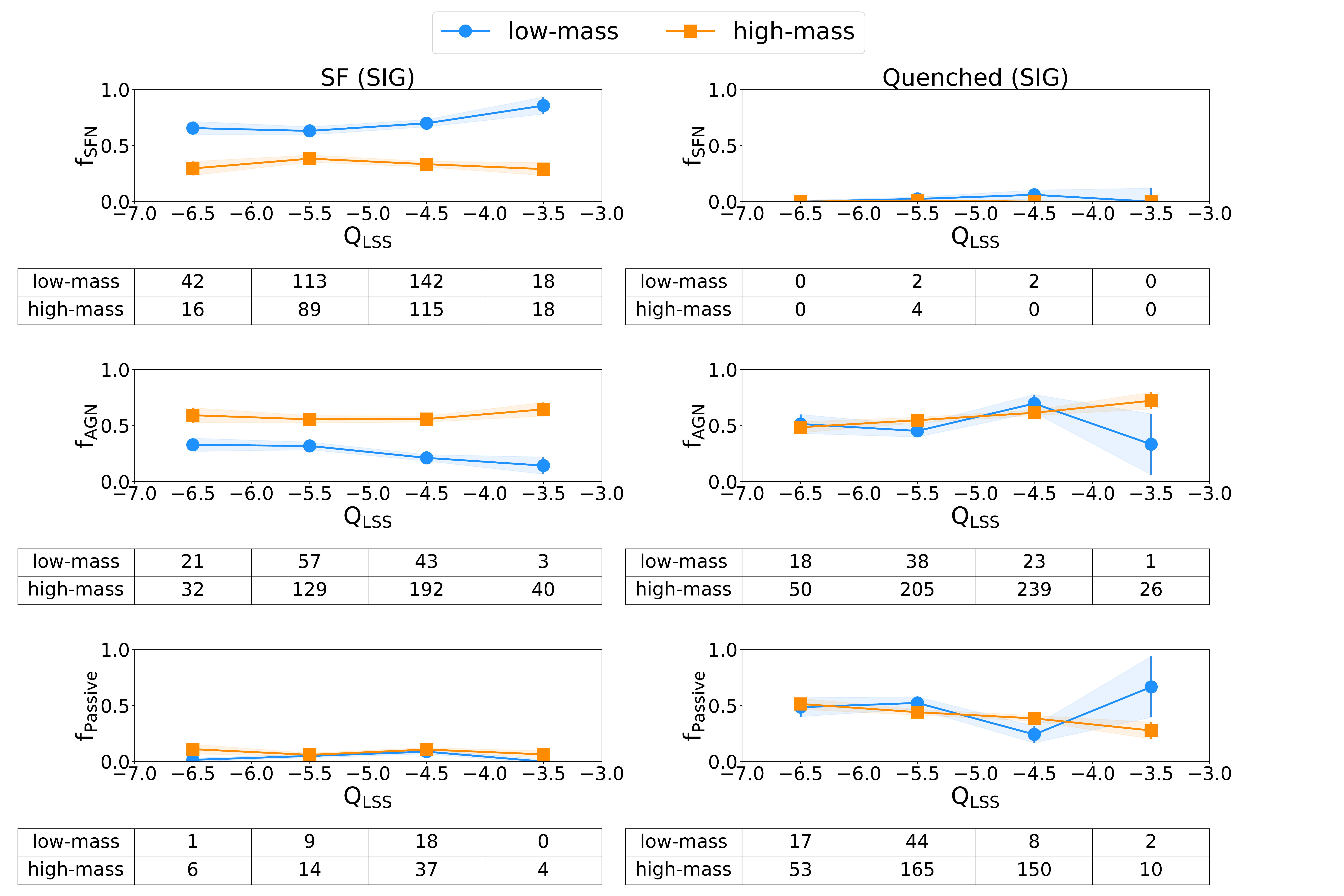}
\caption{Fraction of optical nuclear activity with respect to the $Q_{\rm LSS}$ environmental parameter according to their sSFR status. Low-mass galaxies ($10.0~\leq~\rm{log}(M_\star)~<~10.5~[M_\odot]$) are represented by cyan circles, and high-mass galaxies ($10.5~\leq~\rm{log}(M_\star)~\leq~11.4~[M_\odot]$) are represented by orange squares. The fraction of SFN, optical AGN, and passive star forming SIG galaxies ($N_T$ = 577, 541, and 91, respectively) is represented from top to bottom in the left panels, and for quenched SIG galaxies ($N_T$ = 8, 611, and 468, respectively) in the right panels. The number of galaxies in each $Q_{\rm LSS}$ bin is shown in tables for each sample at the bottom of each panel. Error bars are given considering binomial distribution.}
\label{Fig:App105}
\end{figure*}

\begin{figure*} 
\centering
\includegraphics[width=\textwidth]{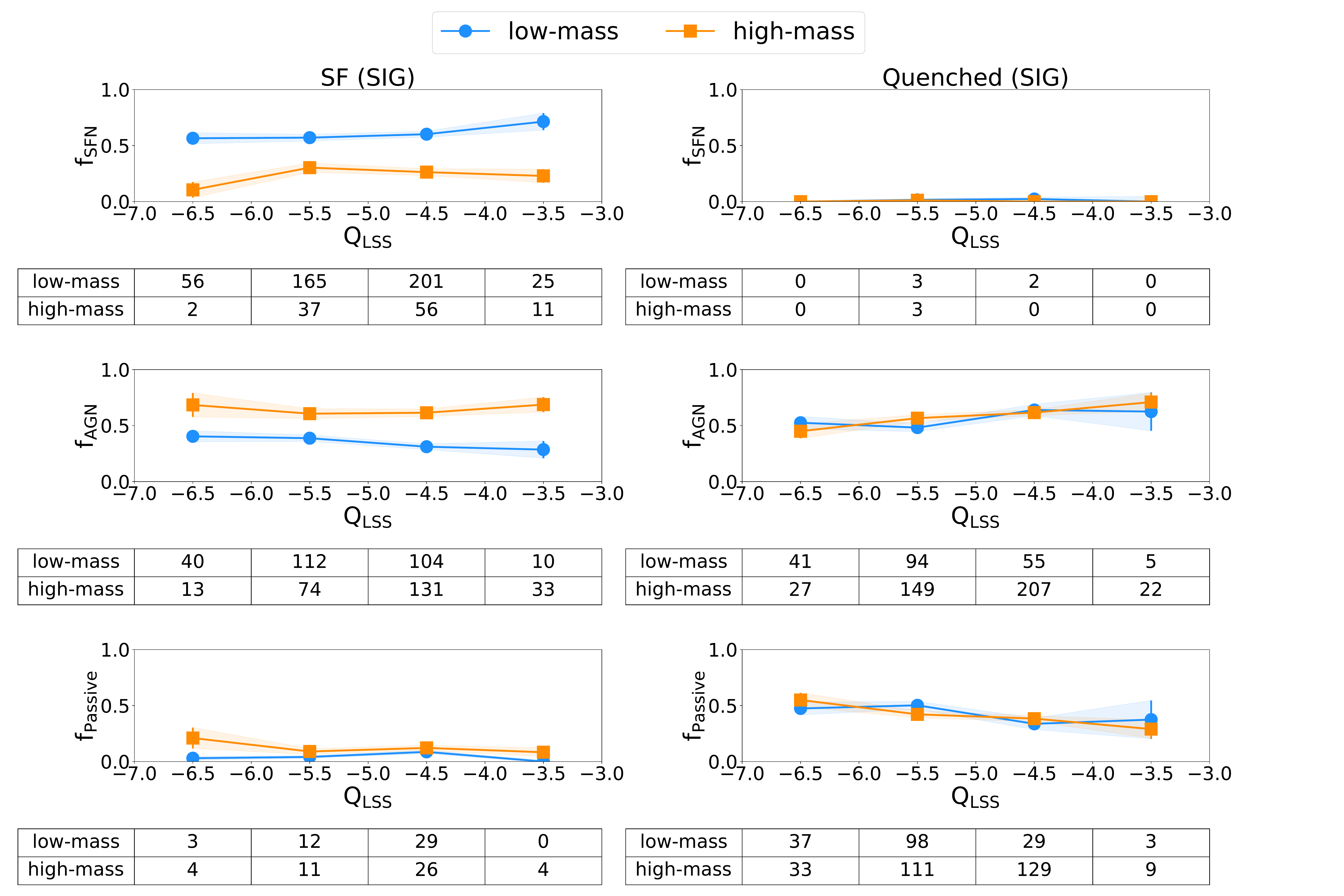}
\caption{Fraction of optical nuclear activity with respect to the $Q_{\rm LSS}$ environmental parameter according to their sSFR status. Low-mass galaxies ($10.0~\leq~\rm{log}(M_\star)~<~10.7~[M_\odot]$) are represented by cyan circles, and high-mass galaxies ($10.7~\leq~\rm{log}(M_\star)~\leq~11.4~[M_\odot]$) are represented by orange squares. The fraction of SFN, optical AGN, and passive star forming SIG galaxies ($N_T$ = 577, 541, and 91, respectively) is represented from top to bottom in the left panels, and for quenched SIG galaxies ($N_T$ = 8, 611, and 468, respectively) in the right panels. The number of galaxies in each $Q_{\rm LSS}$ bin is shown in tables for each sample at the bottom of each panel. Error bars are given considering binomial distribution.}
\label{Fig:App107}
\end{figure*}

Figures~\ref{Fig:App105} and \ref{Fig:App107} are similar to Fig.~\ref{Fig:SIG_LSS_status_106} but using a stellar mass cut $\rm{log}(M_\star)~=~10.5~[M_\odot]$ and $\rm{log}(M_\star)~=~10.7~[M_\odot]$, respectively. We observe that the trends of the fractions of AGN for star-forming and quenched galaxies are somewhat sensitive to the stellar mass cut. In particular, using a mass cut $\rm{log} (M_\star)~=~10.5~[M_\odot]$ the observed trends in low-mass star-forming galaxies are steeper. We also find a separation of the fraction of AGN for the highest bin of $Q_{\rm LSS}$ in quenched galaxies, where the fraction of AGN in low-mass galaxies is lower than the fraction of AGN in high-mass  galaxies. Since this observed separation disappear when selecting $\rm{log} (M_\star)~=~10.6~[M_\odot]$ and $\rm{log} (M_\star)~=~10.7~[M_\odot]$, it may be an artefact of the low number of low-mass quenched galaxies in that bin. Similarly on the fraction of passive quenched galaxies. We do not observe any difference of the trends in terms of morphology and colour, we therefore do not show the corresponding figures.

\end{appendix}

\end{document}